\documentclass[graybox]{svmult}

\usepackage{tikz}
\usepackage{type1cm}        
\usepackage{makeidx}         
\usepackage{graphicx}        
\usepackage{multicol}        
\usepackage[bottom]{footmisc}

\usepackage{newtxtext}       %
\usepackage{newtxmath}       

\makeindex             

\begin{document}

\title*{Jammed disks of two sizes in a narrow channel}
\author{Dan Liu and Gerhard M{\"{u}}ller}
\institute{Dan Liu,  Department of Physics,
  University of Hartford, West Hartford, CT 06117, USA \\ \email{dliu@hartford.edu} \\
Gerhard M{\"{u}}ller,  Department of Physics,
  University of Rhode Island,
  Kingston RI 02881, USA \\ \email{gmuller@uri.edu}}
%
%
\maketitle

\abstract{A granular-matter model is exactly solved, where disks of two sizes and weights in alternating sequence are confined to a narrow channel. 
The axis of the channel is horizontal and its plane vertical.
Disk sizes and channel width are such that under jamming no disks remain loose and all disks touch one wall.
Jammed microstates are characterized via statistically interacting particles constructed out of two-disk tiles.
Jammed macrostates depend on measures of expansion work, gravitational potential energy, and intensity of random agitations before jamming.
The dependence of configurational entropy on excess volume exhibits a critical point.}

\section*{}
\textbf{Introduction}~ 
The research reported here builds on two previous studies \cite{janac1, janac2} employing the same methodology of exact analysis in the framework of configurational statistics.
They, in turn, were inspired by work on jammed disks in a narrow channel based on different methods of analysis \cite{BS06, AB09, BA11} that yielded intriguing results.  
The focus on disks of two sizes and weights in alternating sequence, for which new exact results are being presented here, is intended to be a first step toward a scenario where the channel contains such disks in a random sequence -- a realistic goal in the framework of the same methodology. \\

\noindent \textbf{Geometry}~ 
Disks of two sizes with diameters $\sigma_\mathrm{L}\geq\sigma_\mathrm{S}$ in alternating sequence are being jammed in a channel of width $H$. 
The following two conditions guarantee that (i) the disk sequence remains invariant before jamming, (ii) all jammed disks have wall contact, and (iii) jamming leaves no disks loose:
\begin{equation}\label{eq:1}  
\frac{1}{4}<\frac{\sigma_\mathrm{S}}{\sigma_\mathrm{L}}\leq 1, \qquad
 1<\frac{H}{\sigma_\mathrm{L}}<
\frac{1}{2}\left[1+\frac{\sigma_\mathrm{S}}{\sigma_\mathrm{L}} +
\sqrt{\left(1+\frac{\sigma_\mathrm{S}}{\sigma_\mathrm{L}}\right)^2-1}\;\right],
\end{equation}
All jammed microstates can be assembled from 8 tiles composed of two disks with one disk overlapping (Table~\ref{tab:t1}).
Adding a tile to an already existing string must satisfy the successor rule to maintain mechanical stability under jamming forces.

\begin{table}[htb]
  \caption{Distinct tiles that constitute jammed microstates of disk sequences subject to the conditions (\ref{eq:1}). Mechanical stability rule: \textsf{v} must be followed by \textsf{w} or 2 etc. Motifs pertain to $\sigma_\mathrm{L}=2$, $\sigma_\mathrm{S}=1.4$, $H=2.5$. Volume of tiles: $V_\mathrm{c}=\frac{1}{2}(\sigma_\mathrm{L}+\sigma_\mathrm{S})
+\sqrt{H(\sigma_\mathrm{L}+\sigma_\mathrm{S}-H)}$, $V_\mathrm{f}=\frac{1}{2}(\sigma_\mathrm{L}+\sigma_\mathrm{S})
+\sqrt{\sigma_\mathrm{L}\sigma_\mathrm{S}}$ (assuming unit cross section of channel).}\label{tab:t1}
\begin{center}
\begin{tabular}{cccc|cccc|cccc|cccc} \hline\hline
motif & ~ID~ & rule & vol.~ & ~motif~ & ~ID~ & rule & vol.  &
motif & ~ID~ & rule & vol.~ & ~motif~ & ~ID~ & rule & vol. \\ \hline \rule[-2mm]{0mm}{8mm}
\begin{tikzpicture} [scale=0.2]
\draw (0.0,0.0) -- (3.2,0.0) -- (3.2,2.5) -- (0.0,2.5) -- (0,0);
\filldraw [fill=gray, draw=black] (1,1) circle (1.0);
\filldraw [fill=gray, draw=black] (2.5,1.8) circle (0.7);
\end{tikzpicture}
& \textsf{v} & \textsf{w}, 2 & $V_\mathrm{c}$ &   
\begin{tikzpicture} [scale=0.2]
\draw (0.0,0.0) -- (3.37,0.0) -- (3.37,2.5) -- (0.0,2.5) -- (0,0);
\filldraw [fill=gray, draw=black] (1.0,1.0) circle (1.0);
\filldraw [fill=gray, draw=black] (2.67,0.7) circle (0.7);
\end{tikzpicture}
& 1 & 3, 4 & $V_\mathrm{f}$ & 
\begin{tikzpicture} [scale=0.2]
\draw (0.0,0.0) -- (3.37,0.0) -- (3.37,2.5) -- (0.0,2.5) -- (0,0);
\filldraw [fill=gray, draw=black] (0.7,0.7) circle (0.7);
\filldraw [fill=gray, draw=black] (2.37,1.0) circle (1.0);
\end{tikzpicture}
& 3 & \textsf{v} & $V_\mathrm{f}$ &   
\begin{tikzpicture} [scale=0.2]
\draw (0.0,0.0) -- (3.2,0.0) -- (3.2,2.5) -- (0.0,2.5) -- (0,0);
\filldraw [fill=gray, draw=black] (1,1.5) circle (1.0);
\filldraw [fill=gray, draw=black] (2.5,0.7) circle (0.7);
\end{tikzpicture}
& 5 & 3, 4 & $V_\mathrm{c}$ \\ \rule[-2mm]{0mm}{6mm}

\begin{tikzpicture} [scale=0.2]
\draw (0.0,0.0) -- (3.2,0.0) -- (3.2,2.5) -- (0.0,2.5) -- (0,0);
\filldraw [fill=gray, draw=black] (0.7,1.8) circle (0.7);
\filldraw [fill=gray, draw=black] (2.2,1.0) circle (1.0);
\end{tikzpicture}
& \textsf{w} & \textsf{v}, 1 & $V_\mathrm{c}$ &  
\begin{tikzpicture} [scale=0.2]
\draw (0.0,0.0) -- (3.37,0.0) -- (3.37,2.5) -- (0.0,2.5) -- (0,0);
\filldraw [fill=gray, draw=black] (0.7,1.8) circle (0.7);
\filldraw [fill=gray, draw=black] (2.37,1.5) circle (1.0);
\end{tikzpicture}
& 2 & 5 & $V_\mathrm{f}$ &
\begin{tikzpicture} [scale=0.2]
\draw (0.0,0.0) -- (3.2,0.0) -- (3.2,2.5) -- (0.0,2.5) -- (0,0);
\filldraw [fill=gray, draw=black] (0.7,0.7) circle (0.7);
\filldraw [fill=gray, draw=black] (2.2,1.5) circle (1.0);
\end{tikzpicture}
& 4 & 5, 6 & $V_\mathrm{c}$ &  
\begin{tikzpicture} [scale=0.2]
\draw (0.0,0.0) -- (3.37,0.0) -- (3.37,2.5) -- (0.0,2.5) -- (0,0);
\filldraw [fill=gray, draw=black] (1.0,1.5) circle (1.0);
\filldraw [fill=gray, draw=black] (2.67,1.8) circle (0.7);
\end{tikzpicture}
& 6 & \textsf{w}, 2 & $V_\mathrm{f}$ \\ \hline\hline
\end{tabular}
\end{center}
\end{table}

Under mild assumptions the microstate of minimum volume with $N$ (large/small) disk pairs is composed of an alternating sequence,
\textsf{vwvw$\cdots$ v}, of just two tiles.
We declare it to be the reference state for statistically interacting particles in this application.
All other (jammed) microstates can be generated by the activation of quasi-particles from this reference state. 
We have identified $M=5$ species of particles that serve this purpose (Table~\ref{tab:t3}).
Adopting the taxonomy of Ref.~\cite{copic} we distinguish between the categories of hosts and tags.

\begin{table}[htb]
  \caption{Five species of quasi-particles. The hosts $m=1,\ldots,4$ modify the reference state whereas the tag $m=5$ modifies any one of the hosts. The ID lists the overlapping tiles involved. 
The activation energy $\epsilon_m$ is relevant before jamming and the excess volume $\Delta V_m$ after jamming.}\label{tab:t3}
\begin{center}
\begin{tabular}{lllll|lllll} \hline\hline \rule[-2mm]{0mm}{6mm}
ID & ~$m$ & cat. & ~$\Delta V_m$ & $\epsilon_m$ &
~~ID & ~$m$ & cat. & ~$\Delta V_m$ & $\epsilon_m$  \\ \hline \rule[-2mm]{0mm}{6mm}
 \textsf{13} & ~1 & host & ~$2V_\mathrm{t}$ & $2pV_\mathrm{t}-\gamma_S$ &
 ~~\textsf{2546} & ~4 & host & ~$2V_\mathrm{t}$ & $2pV_\mathrm{t}-\gamma_\mathrm{S}+2\gamma_\mathrm{L}$
\\ \rule[-2mm]{0mm}{5mm}
\textsf{146} & ~2 & host & ~$2V_\mathrm{t}$ & $2pV_\mathrm{t}-\gamma_\mathrm{S}+\gamma_\mathrm{L}$ &  ~~\textsf{45} & ~5 & tag & $0$ & $\gamma_\mathrm{L}-\gamma_\mathrm{S}$
\\ \rule[-2mm]{0mm}{5mm}
\textsf{253} & ~3 & host & ~$2V_\mathrm{t}$ & $2pV_\mathrm{t}-\gamma_\mathrm{S}+\gamma_\mathrm{L}$
\\ \hline\hline
\end{tabular}
\end{center}
\end{table}

Particles from species $m=1,\ldots,4$ can be placed directly into the reference state (pseudo-vacuum), meaning that it is possible to add a tile \textsf{v} or \textsf{w} to the left or to the right as follows: \textsf{w13v}, ~\textsf{w146w}, ~\textsf{v253v}, ~\textsf{v2546w}.
Particles from species $m=5$ are parasitic in the sense that they can only be placed inside a particle from species $m=1,\ldots,4$, at exactly one position and with one disk overlapping as follow:
\textsf{1|45|3}, ~\textsf{1|45|46}, ~\textsf{25|45|3}, ~\textsf{25|45|46}.
The number of tag particles that can be added inside the same host is only limited by the size of the number $N$ of disk pairs in the system.
For example, two tags inside the first host reads $\mathsf{1|4545|3}$.
The minimum number of tiles \textsf{v} or \textsf{w} between two host can be two as in \textsf{13vw146}, one as in \textsf{13v13}, or zero as in \textsf{146253}.\\

\noindent \textbf{Energetics}~ 
The activation of every host particle extends the total volume after jamming by the amount $2V_\mathrm{t}$, where $V_\mathrm{t}\doteq V_\mathrm{f}-V_\mathrm{c}$ (see Tables \ref{tab:t1} and \ref{tab:t3}).
Placing a tag does not change the volume.
The activation energy $\epsilon_m$ assigned to a particle from species $m$ pertains to the state of random agitation before jamming.
It consists of work against the ambient pressure exerted by pistons and gravitational potential energy, all relative to the reference state. 

We are free to choose the mass density of small and large disks. Therefore, the gravitational potential energies $\gamma_\mathrm{L}$ and $\gamma_\mathrm{S}$ are independent parameters as is the expansion work $2pV_\mathrm{t}$.
A fourth parameter is the intensity $T_\mathrm{k}$ of random agitations.
The jamming protocol is explained in \cite{janac1}.
All results coming out of configurational statistics will only depend on three (dimensionless) ratios of four energy parameters:
\begin{equation}\label{eq:4} 
\beta\doteq\frac{2pV_\mathrm{t}}{T_\mathrm{k}},\quad 
\Gamma_\mathrm{L}\doteq\frac{\gamma_\mathrm{L}}{2pV_\mathrm{t}},\quad
\Gamma_\mathrm{S}\doteq\frac{\gamma_\mathrm{S}}{2pV_\mathrm{t}}.
\end{equation}\\

\noindent \textbf{Combinatorics}~ 
The quasi-particles identified in Table~\ref{tab:t3} are statistically interacting in the sense that activating one particle affects the number $d_m$ of open slots for the activation of further particles from each species.
This interaction can be accounted for by a multiplicity expression for jammed microstates involving a generalized Pauli principle \cite{Hald91a} in the form, \cite{copic,Wu94,Isak94},
\begin{equation}\label{eq:5}
W(\{N_m\})=\prod_{m=1}^M\left(\begin{array}{c}d_m+N_m-1 \\ N_m\end{array}\right), 
\quad d_m =A_m-\sum_{m'=1}^M g_{mm'}(N_{m'}-\delta_{mm'}),
\end{equation}
with capacity constants $A_m$ and statistical interaction coefficients $g_{mm'}$ as listed in Table~\ref{tab:t4}, and where $N_m$ is the number of activated particles from species $m$.

\begin{table}[b]
  \caption{Capacity constants and statistical interaction coefficients for each particle species.}\label{tab:t4}
\begin{center}
\begin{tabular}{c|c} \hline\hline
$m$~~ & ~~$A_m$  \\ \hline
1~~ & ~~$N-2$ \\
2~~ & ~~$N-3$ \\
3~~ & ~~$N-2$ \\
4~~ & ~~$N-3$ \\
5~~ & ~~$0$ \\ \hline\hline 
\end{tabular} \hspace{5mm}
\begin{tabular}{c|rrrrr} \hline\hline 
$g_{mm'}$ & $1$ & $2$ & $3$&4&5 \\ \hline 
$1$ & 2 & 2 & 1 & 2 &1\\ 
$2$ & 1 & 2 & 1& 1 &1\\ 
$3$ & 2 & 2& 2&2&1\\
$4$ & 1 & 2 & 1&2&1\\
$5$ & ~$-1$ & $-1$ & $-1$&$-1$&~~0\\ \hline\hline
\end{tabular}
\end{center}
\end{table}

The initial capacity for hosts grows linearly with the number of disks.
It is zero for tags, which can only be activated inside hosts.
Activating a host $(m'=1,\ldots,4)$ removes one or two slots for activating a further host $(m=1,\ldots,4)$ but adds one slot for activating a tag $(m=5)$. 
Activating a tag $(m'=5)$ removes one slot for activating hosts $(m=1,\ldots,4)$ but leaves the number of slots for activating a further tag $(m=5)$ invariant.\\

\noindent \textbf{Statistical mechanics}~ 
We can express the excess volume and the entropy as functions of the average particle content $\{\langle N_m\rangle\}$ of a jammed macrostate as follows \cite{Isak94}:
\begin{subequations}\label{eq:7}
\begin{equation} \label{eq:7a}
V-V_\mathrm{ref}=\sum_{m=1}^M\langle N_m\rangle \Delta V_m,\qquad 
Y_m \doteq A_m-\sum_{m'=1}^Mg_{mm'} \langle N_{m'}\rangle.
\end{equation}
\begin{equation}\label{eq:7b}
S = k_B\sum_{m=1}^M\Big[\big(\langle N_{m}\rangle
+Y_m\big)\ln\big(\langle N_m\rangle+Y_m\big) 
-\langle N_m\rangle \ln \langle N_m\rangle -Y_m\ln Y_m\Big].
\end{equation}
\end{subequations}
The average particle numbers are the solutions of the linear equations \cite{Wu94,Isak94},
\begin{equation}\label{eq:9}
w_m \langle N_m\rangle+\sum_{m'=1}^Mg_{mm'} \langle N_{m'}\rangle =A_m.
 \end{equation} 
The $w_m$ are non-negative solutions of the algebraic equations \cite{copic, Wu94,Isak94, Anghel, LVP+08, picnnn, pichs},
\begin{equation}\label{eq:10} 
e^{\epsilon_m/T_\mathrm{k}}=(1+w_m)\prod_{m'=1}^M \big(1+w_{m'}^{-1}\big)^{-g_{m'm}}.
\end{equation}
The analytic solution of Eqs.~(\ref{eq:10}), too unwieldy for display, gives us explicit expressions for the scaled excess volume, $\bar{V}\doteq (V-V_\mathrm{ref})/NV_\mathrm{t}$, the scaled entropy, $\bar{S}\doteq S/Nk_\mathrm{B}$, and the particle densities, $\bar{N}_m\doteq\langle N_m\rangle/N$, as functions of $\beta, \Gamma_\mathrm{L}, \Gamma_\mathrm{S}$.\\

\noindent \textbf{Results}~
There is space to present one case, $\Gamma_\mathrm{L}=\Gamma_\mathrm{S}$, at the border between two regimes, large and small disks have equal gravitational potential energy when they touch the same wall.
Only two of the parameters (\ref{eq:4}) are independent.
Increasing $\beta$ means reducing the intensity of random agitations before jamming and increasing $\Gamma_\mathrm{L}$ means increasing the effects of gravity (e.g. by tilting the the plane of the channel from horizontal toward vertical).

\begin{figure}[htt]
  \begin{center}
\includegraphics[width=55mm]{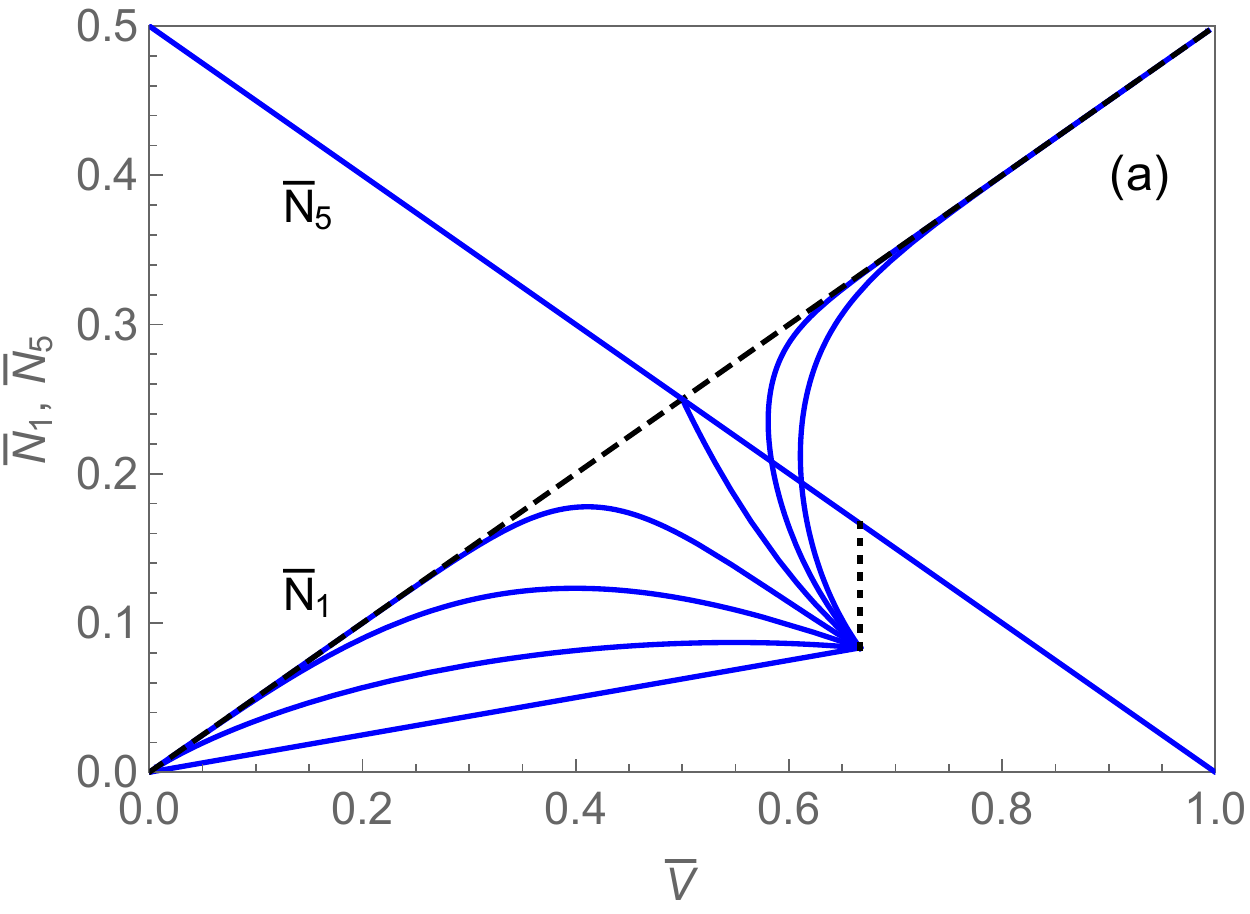}%
\hspace*{3mm}\includegraphics[width=55mm]{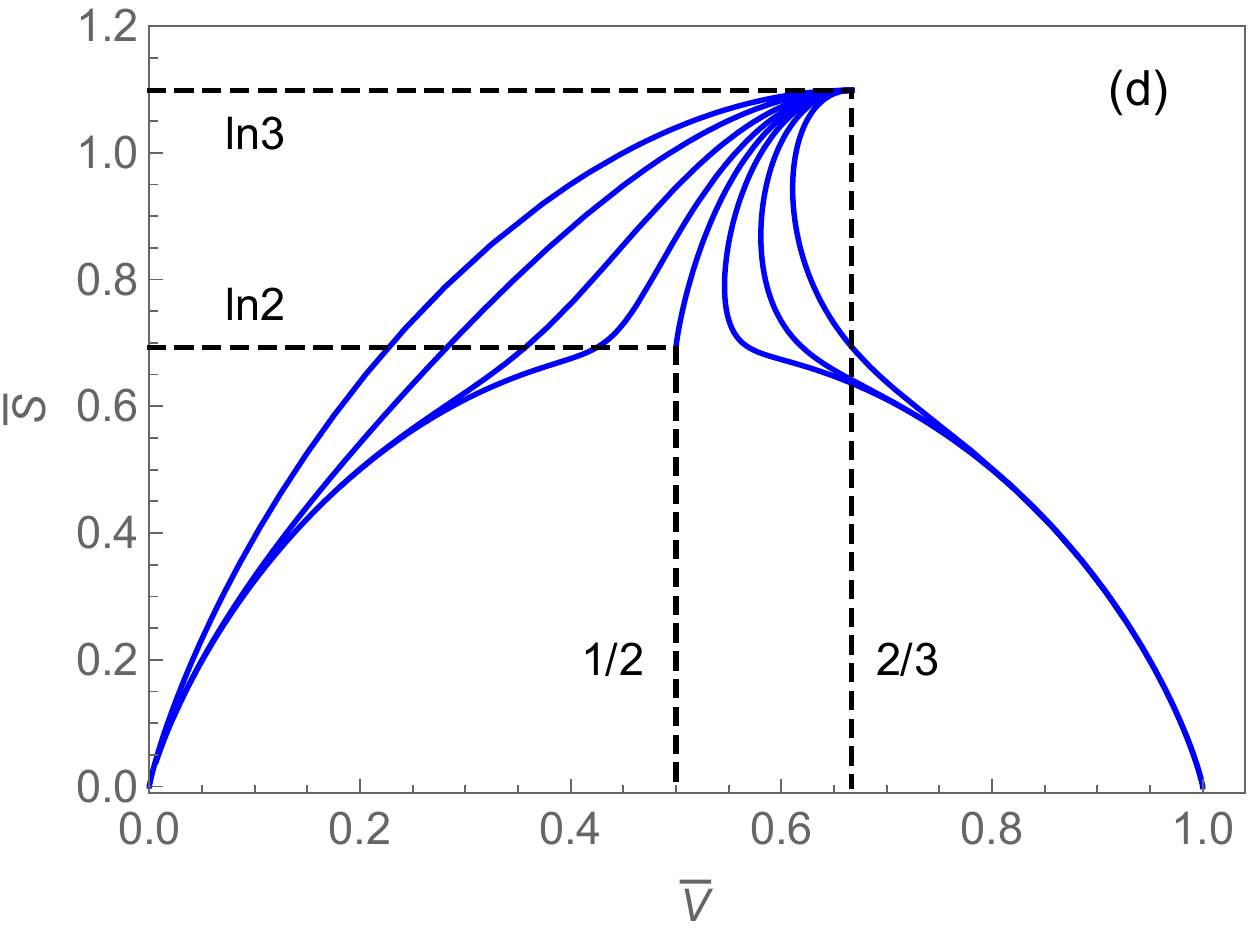}
\includegraphics[width=43mm]{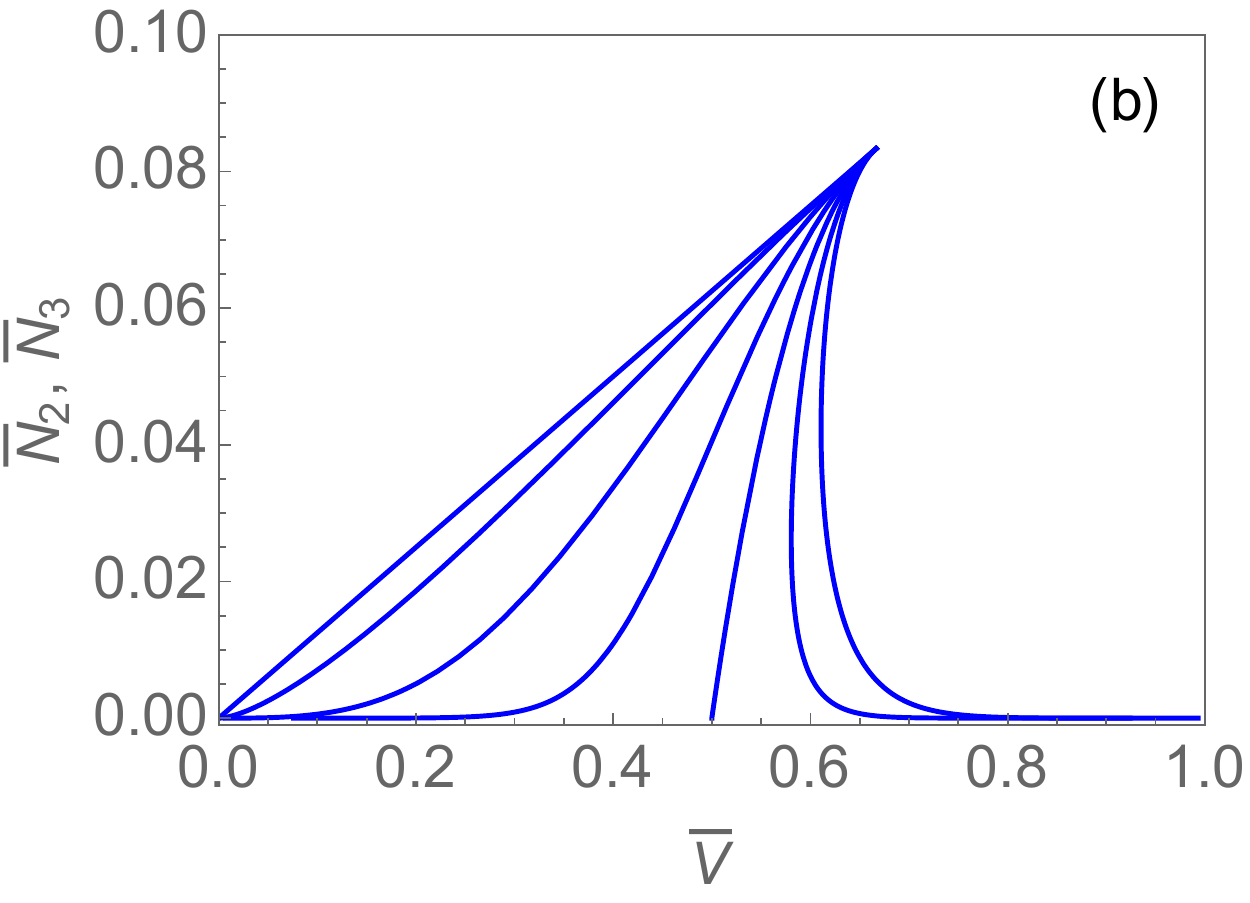}%
\hspace*{3mm}\includegraphics[width=43mm]{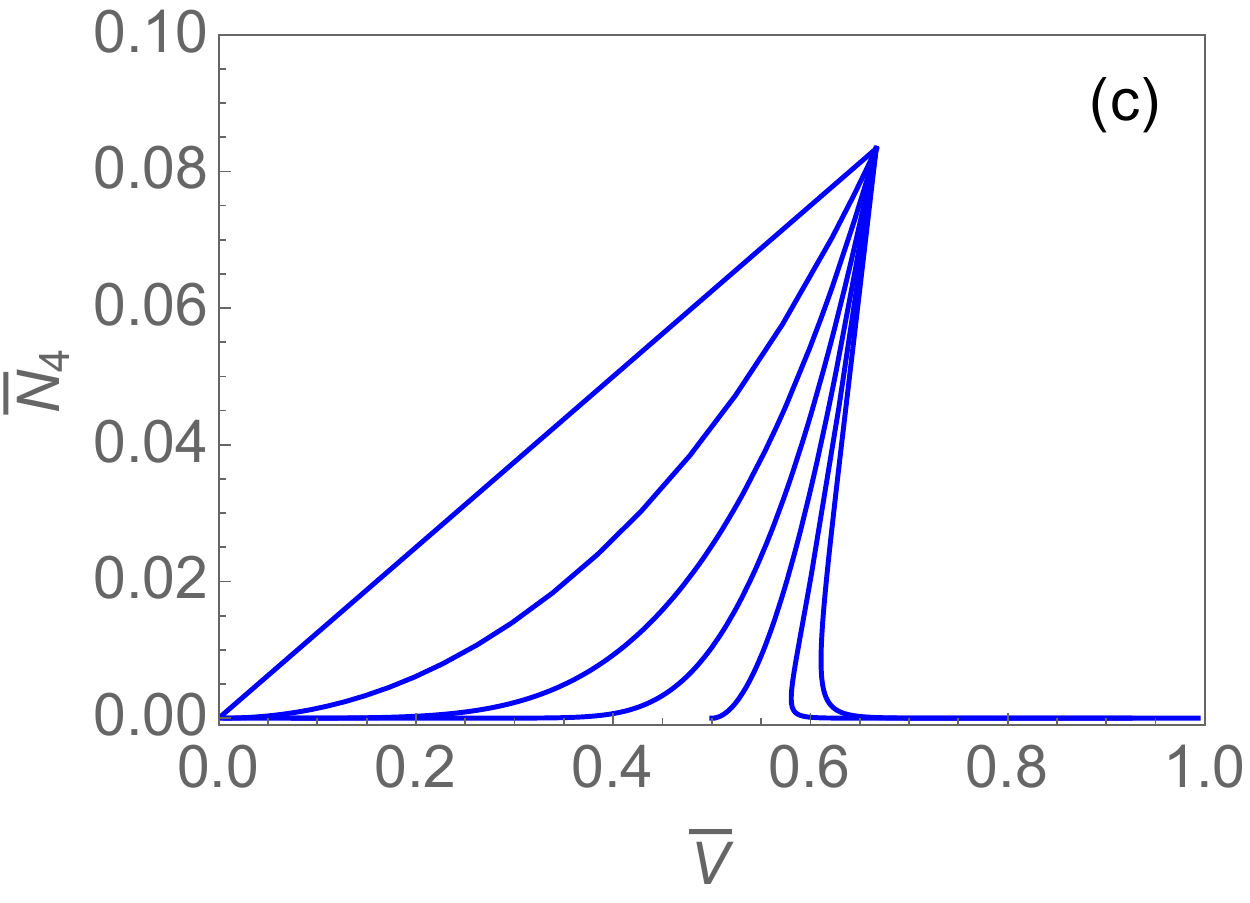}
\end{center}
\caption{Population densities $\bar{N}_m$, $m=1,\ldots,5$ versus excess volume $\bar{V}$ for fixed values 0, 0.5, 0.75, 1.0, 1.25, 1.5 of parameter $\Gamma_\mathrm{L}=\Gamma_\mathrm{S}$ from bottom up in panel (a) and from left to right in panels (b) and (c)]. 
The dotted line connect the points pertaining to $\beta=0$ of $\bar{N}_1$ and $\bar{N}_5$. 
The dashed diagonal represents the sum $\bar{N}_1+\bar{N}_2+\bar{N}_3+\bar{N}_4$.
(d) Entropy $\bar{S}$ versus excess volume $\bar{V}$ for fixed values 0, 0.5, 0.75, 0.9, 1.0, 1.1, 1.25, 1.5 of parameter $\Gamma_\mathrm{L}=\Gamma_\mathrm{S}$ (from left to right).
}
  \label{fig:f3}
\end{figure}

In the high-intensity limit, $\beta\to0$, we have the most disordered macrostate:
\begin{equation}\label{eq:18}
\bar{N}_1=\cdots=\bar{N}_4=\frac{1}{12},\quad \bar{N}_5=\frac{1}{6},\quad 
\bar{V}=\frac{2}{3},\quad \bar{S}=\ln3\qquad (\beta=0).
\end{equation}
It is instructive to plot the the population densities versus volume [Fig.~\ref{fig:f3}(a)-(c)].
For $\bar{N}_1,\ldots,\bar{N}_4$, $\beta=0$ is realized at the points where multiple curves merge. 
For $\bar{N}_5$ that point is at the top of the dotted line.
Increasing $\beta$ toward infinity means moving along any path toward $\bar{V}=0$ if $\Gamma_\mathrm{L}<1$ or toward $\bar{V}=1$ if $\Gamma_\mathrm{L}>1$.
The path stops at $\bar{V}=\frac{1}{2}$ if $\Gamma_\mathrm{L}=1$, which signals criticality.
The following conservation law only holds for the case considered here:
\begin{equation}\label{eq:2} 
\bar{N}_\mathrm{tot}\doteq\sum_{m=1}^5\bar{N}_m=\frac{1}{2}\qquad 
(\Gamma_\mathrm{L}=\Gamma_\mathrm{S}).
\end{equation}
Its validity is illustrated by the two diagonal lines in Fig.~\ref{fig:f3}(a), where the dashed line represents the population density of all hosts combined.
The tags $m=5$ do not contribute to excess volume.
Their numbers depend on $\bar{V}$ nevertheless, via the conservation law (\ref{eq:2}).
Any increase in volume caused by the activation of one or the other host necessarily crowds out one tag.

The results show that hosts $m=2,3,4$ only contribute significantly at high intensity.
The three distinct macrostates associated with $\beta=\infty$ are located at the corners or the center of Fig.~\ref{fig:f3}(a).
All have $\bar{N}_2=\bar{N}_3=\bar{N}_4=0$.

\begin{itemize}

\item The state with $\bar{N}_1=0$ and $\bar{N}_5=\frac{1}{2}$ is realized for $\Gamma_\mathrm{L}<1$ and has $\bar{V}=0$, $\bar{S}=0$.
It is a doublet consisting of the reference state $\mathsf{vwvw\cdots vwv}$ and the state $\mathsf{14545\cdots4546}$.
The former contains no particles, $\bar{N}_5=0$, and the latter one host 2 and a macroscopic number of tags inside, amounting to $\bar{N}_5=1$. 

\item The state with $\bar{N}_1=\frac{1}{2}$ and $\bar{N}_5=0$ is realized for $\Gamma_\mathrm{L}>1$ and has $\bar{V}=1$, $\bar{S}=0$.
It is a singlet packed with hosts 1: $\mathsf{13vw13vw\cdots13v}$. Hosts 1  proliferate because they have negative activation energies.

\item The state with $\bar{N}_1=\bar{N}_5=\frac{1}{4}$ is realized for $\Gamma_\mathrm{L}=1$ and has $\bar{V}=\frac{1}{2}$, $\bar{S}=\ln2$.
It is highly degenerate. The hosts 1 are randomly distributed between vacuum tiles with a random number of tags inside. Hosts 1 and tags 5 have zero activation energies whereas the hosts, 2, 3, 4 have positive activation energies.

\end{itemize}

With all this information in place we are ready to interpret a graphical representation of entropy versus excess energy [Fig.~\ref{fig:f3}(d)].
The parameter $\beta$ runs from zero to infinity along each path from the top down.
All paths start at coordinates $\bar{V}=\frac{2}{3}$, $\bar{S}=\ln3$.
Paths for $\Gamma_\mathrm{L}<1$ end at $\bar{V}=0$, $\bar{S}=0$, and paths for $\Gamma_\mathrm{L}>1$ at $\bar{V}=1$, $\bar{S}=0$.
At critical gravity, $\Gamma_\mathrm{L}=1$, both volume and entropy decrease monotonically but end at the critical values $\bar{V}=\frac{1}{2}$, $\bar{S}=\ln2$.

Compact analytic expressions for the curves in Fig.~\ref{fig:f3} pertaining to zero gravity and critical gravity are available.
For $\Gamma_\mathrm{L}=0$ we have
\begin{equation}\label{eq:20}
\bar{N}_1=\cdots=\bar{N}_4=\frac{1}{8}\bar{V},\quad 
\bar{N}_5=\frac{1}{2}\big(1-\bar{V}\big),\quad 
\bar{S}=\big(\bar{V}-1\big) \ln \left(\frac{2}{\bar{V}}-2\right)+\ln
   \left(\frac{2}{\bar{V}}\right),
\end{equation}
where the range of volume is $0\leq\bar{V}\leq\frac{2}{3}$.
For $\Gamma_\mathrm{L}=1$ we have
\begin{subequations}\label{eq:21}
\begin{align}\label{eq:21a}
& \bar{N}_1=\frac{(\bar{V}-1)^2}{2\bar{V}},\quad 
\bar{N}_2=\bar{N}_3=\frac{3}{2}-\bar{V}-\frac{1}{2\bar{V}},\quad
\bar{N}_4=2\bar{V}+\frac{1}{2\bar{V}}-2, \\ \label{eq:21b}
& \bar{N}_5=\frac{1}{2}\big(1-\bar{V}\big), \quad 
 \bar{S}=2 (\bar{V}-1) \ln \left(\frac{1-\bar{V}}{2
   \bar{V}-1}\right)-\ln (2 \bar{V}-1),
\end{align}
\end{subequations}
across the more restricted range $\frac{1}{2}\leq\bar{V}\leq\frac{2}{3}$ of volume.\\

\noindent \textbf{Outlook}~
Qualitatively different jamming patterns pertain to the regimes ${\Gamma_\mathrm{L}<\Gamma_\mathrm{S}}$ of light large disks and $\Gamma_\mathrm{L}>\Gamma_\mathrm{S}$ of heavy large disks.
Yet different jamming patterns are expected when the analysis is generalized to periodic and aperiodic sequences including random sequences of large and small disks (with modified jamming protocols).

\end{document}